\documentclass[pra,showpacs]{revtex4}

\usepackage[dvips]{graphics}
\usepackage{amsmath, amsthm, amssymb}
\usepackage{graphicx}
\usepackage{dcolumn}
\usepackage{bm}

\newcommand{\de}[1]{\left( #1 \right)}

\newcommand{\DE}[1]{\left\{ #1 \right\}}

\newcommand{\ket}[1]{\left| #1 \right\rangle}
\newcommand{\bra}[1]{\left\langle #1 \right|}

\newcommand{\tr}{\mathrm{Tr}}

\newcommand{\eg}{{\it{e.g.}}}
\newcommand{\ie}{{\it{i.e.}}}
\newcommand{\etal}{{\it{et al.}}}

\begin{document}

\author{Daniel Cavalcanti$^{1}$, J. G. Oliveira Jr.$^{1}$, J. G. Peixoto de Faria$^{2}$,
Marcelo O. Terra Cunha$^{3,4}$, and Marcelo Fran\c ca Santos$^{1}$}
\email{msantos@fisica.ufmg.br}

\address{$1$ Departamento de F\'{\i}sica - CP 702 - Universidade
Federal de Minas Gerais - 30123-970 - Belo Horizonte - MG - Brazil\\
$2$ Departamento Acad\^emico de Disciplinas B\'asicas - Centro
Federal de Educa\c c\~ao Tecnol\'ogica de Minas Gerais - 30510-000 -
Belo Horizonte - MG - Brazil\\
$3$ The School of Physics and Astronomy, University of Leeds, Leeds
LS2 9JT, UK\\
$4$ Departamento de Matem\'atica - CP 702 - Universidade Federal de
Minas Gerais - 30123-970 - Belo Horizonte - MG - Brazil}

\title{Entanglement versus energy in the entanglement transfer problem}

\begin{abstract}
We study the relation between energy and entanglement in an
entanglement transfer problem. We first analyze the general setup of
two entangled qubits (``$a$'' and ``$b$'') exchanging this entanglement
with two other independent qubits (``$A$'' and ``$B$''). Qubit ``$a$''
(``$b$'') interacts with qubit ``$A$'' (``$B$'') via a spin exchange-like
unitary evolution. A physical realization of this scenario could be
the problem of two-level atoms transferring entanglement to resonant
cavities via independent Jaynes-Cummings interactions. We study the
dynamics of entanglement and energy for the second pair of qubits
(tracing out the originally entangled ones) and show that these
quantities are closely related. For example, the allowed quantum
states occupy a restricted area in a phase diagram entanglement vs.
energy. Moreover the curve which bounds this area is exactly the one
followed if both interactions are equal and the entire four qubit
system is isolated. We also consider the case when the target pair
of qubits is subjected to losses and can spontaneously decay.
\end{abstract}

\pacs{03.67.-a, 03.67.Mn, 03.65.Yz}

\maketitle

\section{Introduction}

Entanglement is one of the most studied topics at present. The large
interest for this issue relies mainly on the fact that entangled
systems can be used to perform some tasks more efficiently than
classical objects~\cite{NC}. It is then natural to look for a good
understanding of this resource not only from a purely mathematical
point of view, {\it{i.e.}}, formalizing the theory of entanglement, but also
from a more practical approach, {\it{i.e.}}, studying its role and
manifestations in realistic systems. For example, recent works have
been able to connect entanglement to thermodynamical properties of
macroscopic physical systems~\cite{term}.
In a distinct venue, other works study physical manifestations of
quantum correlations by suitably choosing particular purity and
entanglement quantifiers and restricting allowed quantum states
according to these quantities~\cite{quant}. In these studies,
concepts like \emph{maximally entangled mixed states (MEMS)} are
discussed.
A very recent study also adds energy to entanglement and purity as a
third parameter to characterize certain quantum
states~\cite{Buzek2}. In particular, the authors discuss the
physically allowed states according to the possible values of
entanglement, purity and energy for a system composed of two qubits
or two gaussian states, {\rm and also study the entanglement
transfer between them}.


In the present manuscript, we study the connection between
entanglement and energy that appears naturally in a swapping process
involving two systems of two qubits. In the model investigated, we
consider a simple form of interaction between two pairs of qubits
labeled as $aA$ and $bB$. The system $ab$ is prepared in an
entangled state while the pair $AB$ is prepared in a factorable
state. We analyze the dynamical relations between energy and
entanglement of qubits $AB$ when exchanging energy and coherence
with qubits $ab$. In particular, for any given time $t$, we
calculate the full quantum state of qubits $abAB$ and then we trace
out qubits $ab$ to calculate energy and entanglement of the
remaining pair $AB$. We show that this dynamics yields paths in an
entanglement-energy diagram, and that these paths are contained in a
very restricted region. Moreover, we identify the frontiers of this
region from the general form of the density operator that represents
the state of the subsystem $AB$. We also propose a physical system
to realize such entanglement transfer and investigate how the
dynamics of the entanglement swapping is modified if the $AB$ system
is open and allowed to dissipate energy to an external reservoir. In
some sense, this work is complementary to the sequence \cite{Mauro}
in which the authors study the problem of entanglement transfer from
continuous-variable entangled states to qubits, although in those
works the authors do not pay particular attention to the relation
between Energy and Entanglement.

The paper is organized as follows. In section \ref{PhysScen} we
introduce the general physical system that we will study and the
basic setup from which we will approach it. We also define the
quantities that will be analyzed throughout the paper and finally we
discuss the dynamics of this system. Section \ref{EntEn} is devoted
to study the entanglement and the energy of system $AB$ under a
particular unitary evolution. We then propose a physical
implementation for the studied Hamiltonian, and generalize the time
evolution in section \ref{OpSys} by considering the problem of a
dissipative, non-unitary evolution. In section \ref{Conc} we
conclude by reviewing the main points we have discussed and
suggesting possible extensions of this study.

\section{Physical scenario}\label{PhysScen}

Let us start by describing the system we are interested in. Suppose
a system of four qubits $a$, $b$, $A$, and $B$ interacting via a
spin-exchange like Hamiltonian:
\begin{equation}\label{hamilt1}
H=H_{aA}+H_{bB},
\end{equation}
where
\begin{subequations}\label{HxX}
\begin{equation}\label{HaA}
H_{aA}=\frac{\hbar\omega_{a}}{2}\sigma_{z}^{a}+
\frac{\hbar\omega_{A}}{2}\sigma_{z}^{A}+
g_{aA}(\sigma_{-}^{a}\sigma_{+}^{A}+\sigma_{+}^{a}\sigma_{-}^{A})
\end{equation}
and
\begin{equation}\label{HbB}
H_{bB}=\frac{\hbar\omega_{b}}{2}\sigma_{z}^{b}+
\frac{\hbar\omega_{B}}{2}\sigma_{z}^{B}+
g_{bB}(\sigma_{-}^{b}\sigma_{+}^{B}+\sigma_{+}^{b}\sigma_{-}^{B}).
\end{equation}
\end{subequations}
For each qubit, the relevant Pauli operators are defined by
\begin{subequations}
 \begin{eqnarray}
  \sigma _z &=& \ket{1}\bra{1} - \ket{0}\bra{0},\\
  \sigma _+ &=& \ket{1}\bra{0},\\
  \sigma _- &=& \ket{0}\bra{1},
 \end{eqnarray}
\end{subequations}
 and the interaction operators like $\sigma_{-}^{a}\sigma_{+}^{A}$, for
example, can be viewed as annihilating an excitation of subsystem
$a$ and creating an excitation in subsystem $A$. The constants
$g_{aA}$ and $g_{bB}$ give the strength of the interaction between
these subsystems. One important feature in understanding such
Hamiltonians is that the total number of excitations is a conserved
quantity.
The eigenvectors of \eqref{HaA} (similarly to \eqref{HbB}) are given by:
$\ket{00}_{aA}$,
 with eigenvalue $E_{00}^{aA}=-\hbar\omega$,
$\ket{11}_{aA}$,
 with eigenvalue $E_{11}^{aA}=\hbar\omega$,
and $\ket{\Lambda_{\pm}}_{aA}=(\ket{01}\pm \ket{10})/\sqrt{2}$,
 with eigenvalue $E_{\pm}^{aA}=\pm \hbar g_{aA}/2$, where
$\omega = (\omega_a+\omega_A)/2$.

As the initial state, let us suppose that the entire $abAB$ system
is prepared in the form:
\begin{equation}
\ket{\phi(t=0)}=\ket{\psi(\theta)}_{ab}\otimes\ket{00}_{AB}.
\label{state0}
\end{equation}
where
$\ket{\psi(\theta)}_{ab}=\sin\theta\ket{01}+\cos\theta\ket{10}$,
which means that subsystem $AB$ is prepared in its ground state and
subsystem $ab$ is usually prepared in some entangled state with one
excitation (except if $\theta = n\frac{\pi}{2}$, $n \in \mathbb{Z}$,
when the state is factorable). Note that this initial state is pure
and it is chosen so that the bipartition $ab\otimes AB$ does not
present any initial entanglement. From now on, we will study the
time evolution of this initial state when subjected to
Hamiltonian~\eqref{hamilt1} for different coupling constants
$g_{aA}$ and $g_{bB}$. We will concentrate our analysis in the
subsystem $AB$ by tracing out the degrees of freedom of systems $a$
and $b$. Another simplifying assumption we made is to consider the
complete resonance condition $\omega _a=\omega _A=\omega _b=\omega
_B=\omega$.

A special case of this dynamics happens when $g_{aA}=g_{bB}=g$, in
which case state~\eqref{state0} evolves into state
\begin{equation}\label{globalstate}
\ket{\phi(t)}=\cos(g
t)\ket{\psi}_{ab}\otimes\ket{00}_{AB}-i\sin(gt)\ket{00}_{ab}\otimes\ket{\psi}_{AB}.
\end{equation}
Note that in this simple case, for $t=n\frac{\pi}{2g}$, with $n$
odd, the subsystems exchange their states, the entanglement
initially present in subsystem $ab$ is completely transferred to
subsystem $AB$ and with respect to the bipartition $ab \otimes AB$
the state becomes again separable. However, for $t \neq
n\frac{\pi}{2g}$, the whole system is entangled (as long as $\theta
\neq n\frac{\pi}{2}$) and subsystem $AB$ will be in some mixed
state.

In a more general situation (different coupling constants),
state~\eqref{state0} will evolve into:
\begin{eqnarray}
\ket{\phi(t)}&=&\cos\theta [\cos(g_{aA}
t)\ket{1000}-i\sin(g_{aA} t)\ket{0010} ]\nonumber \\
&+&\sin\theta[\cos(g_{bB} t)\ket{0100}-i\sin(g_{bB}
t)\ket{0001}]\label{phi(t)}.
\end{eqnarray}
Note that for generic times $t$, state~\eqref{phi(t)} presents,
again, multipartite entanglement among all its individual components
($a$, $b$, $A$ and $B$). Studying this multipartite entanglement may
also prove intriguing and enlightening. However this is not the
purpose of this manuscript where, as mentioned above, we will
concentrate our analysis in the subsystem $AB$.

Our goal is to investigate the relation between energy and
entanglement in subsystem $AB$ as a function of coupling constants
and time. In order to study entanglement we will use the
\emph{negativity} ($N$) which can be defined for two qubits as two
times the modulus of the negative eigenvalue of the partial
transposition of the state $\rho$, $\rho^{T_A}$ \cite{negat}, if it exists.
For short:
\begin{equation}
N(\rho)=2\max\{0,-\lambda_{min}\},
\end{equation}
where $\lambda_{min}$ is the lowest eigenvalue of $\rho^{T_A}$. Our
choice is motivated by the facts that the Negativity is easy to
calculate and provides full entanglement information for a two-qubit
system. For the energy of subsystem $AB$, $U$, we will consider the
mean value of the relevant restriction of the free Hamiltonian:
\begin{equation}\label{U}
U=\tr\DE{\rho H_{AB}},
\end{equation}
where
\begin{equation}
H_{AB}=\frac{\hbar\omega}{2}\de{\sigma_{z}^{A}+\sigma_{z}^{B}}.
\end{equation}


\section{Entanglement and Energy}\label{EntEn}

After tracing out the degrees of freedom of systems $a$ and $b$ in
the global quantum state \eqref{phi(t)}, the reduced state for the
pair $AB$ is described by:
\begin{equation}\label{stateAB}
\rho_{AB}=\left(%
\begin{array}{cccc}
  a & 0 & 0 & 0 \\
  0 & b & d & 0 \\
  0 & d^* & c & 0 \\
  0 & 0 & 0 & 0 \\
\end{array}%
\right),
\end{equation}
where $a+b+c=1$ (from the normalization of $\rho_{AB}$), with $a$,
$b$, $c$ and $d$ given by the following functions of the coupling
constants and time:
\begin{subequations}
\begin{eqnarray}
a&=&\cos^2\theta\cos^2(g_{aA}t)+\sin^2\theta\cos^2(g_{bB}t),\label{aunit}\\
b&=&\sin^2\theta\sin^2(g_{bB}t),\\
c&=&\cos^2\theta\sin^2(g_{aA}t),\\
d&=&\cos\theta\sin\theta\sin(g_{aA}t)\sin(g_{bB}t).
\end{eqnarray}
\label{elements}
\end{subequations}
Following Eq.~\eqref{U}, the
energy of state \eqref{stateAB} is:
\begin{equation}\label{GenU}
U=-a,
\end{equation}
which, by means of Eqs. \eqref{elements}, becomes:
\begin{equation}
U=-\cos^2\theta\cos^2(g_{aA}t)-\sin^2\theta\cos^2(g_{bB}t).
\end{equation}
Note that $-1\leq U\leq 0$, which means that there is at most one
excitation on the $AB$ system. This is expected since the chosen
initial state contains only one excitation for the entire $abAB$
system and this set of qubits is isolated, i.e. it cannot be excited
by external sources. A simple calculation gives for the entanglement
(negativity) of state \eqref{stateAB}
\begin{equation}\label{negat}
N=\sqrt{a^2+4d^2}-a.
\end{equation}

\subsection{Entanglement and energy versus time}
\begin{figure}[h]\centering
\begin{tabular}{ccc}
\rotatebox{270}{\includegraphics[scale=0.45]{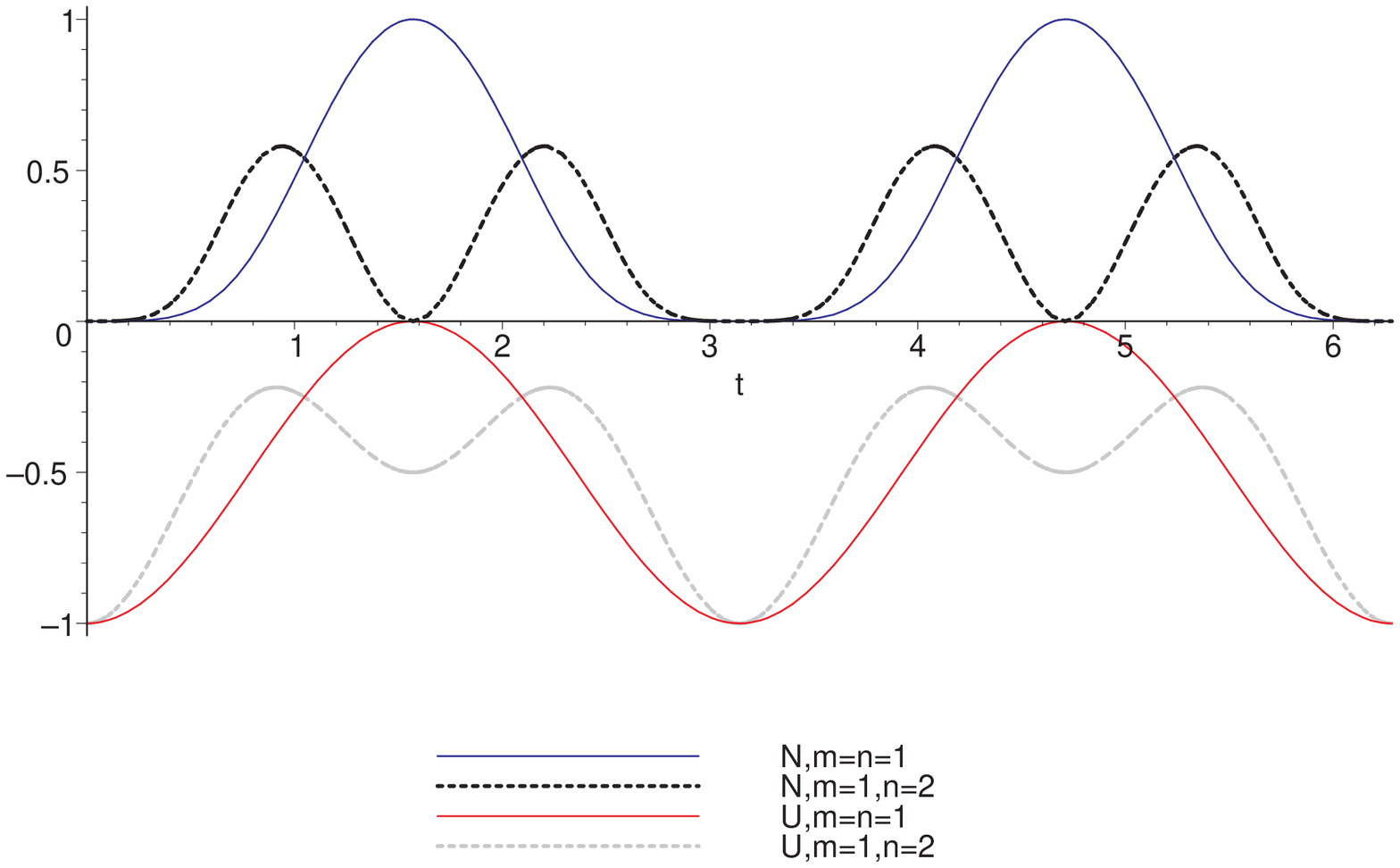}} \\
\rotatebox{270}{\includegraphics[scale=0.45]{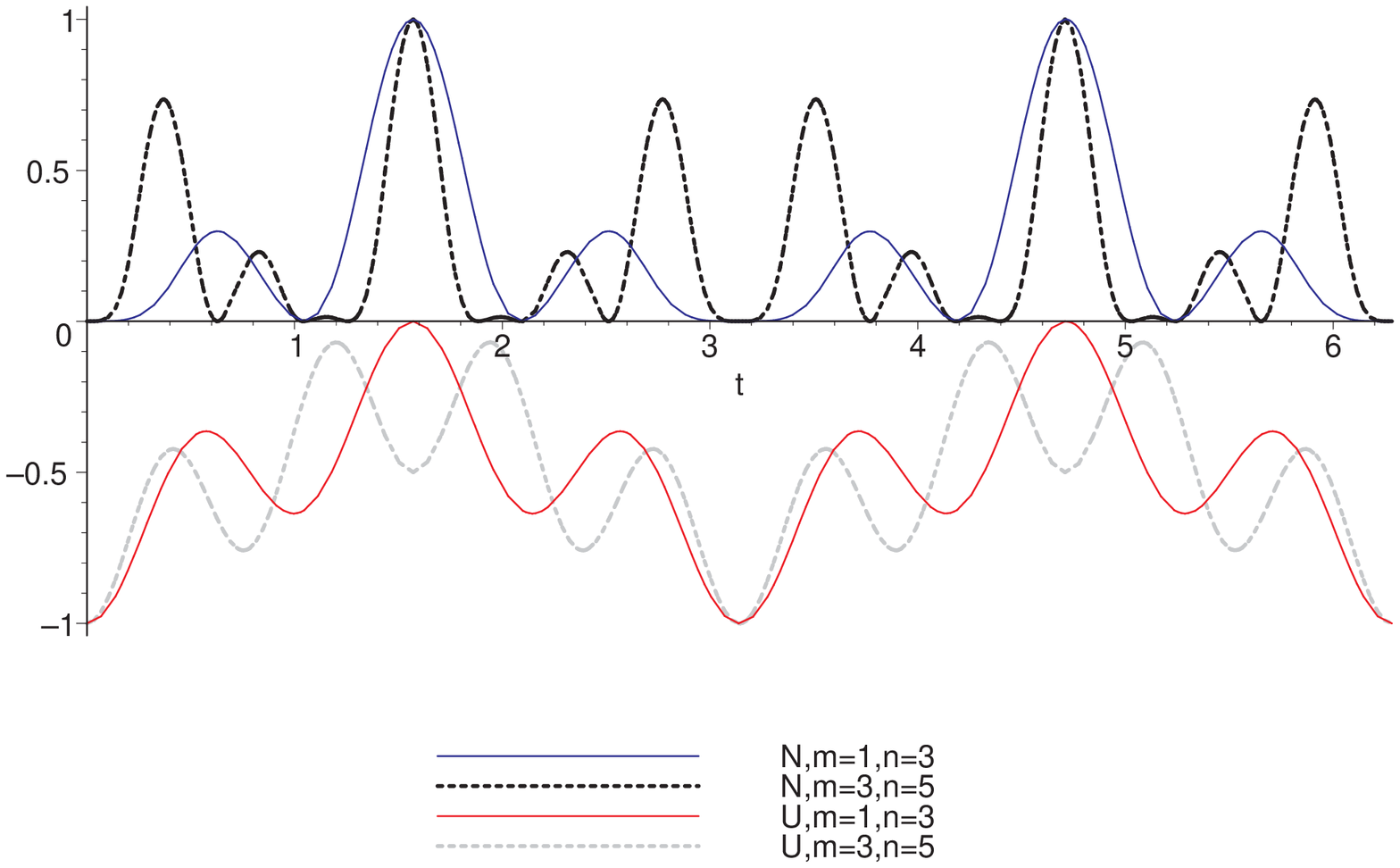}} \\
\rotatebox{270}{\includegraphics[scale=0.45]{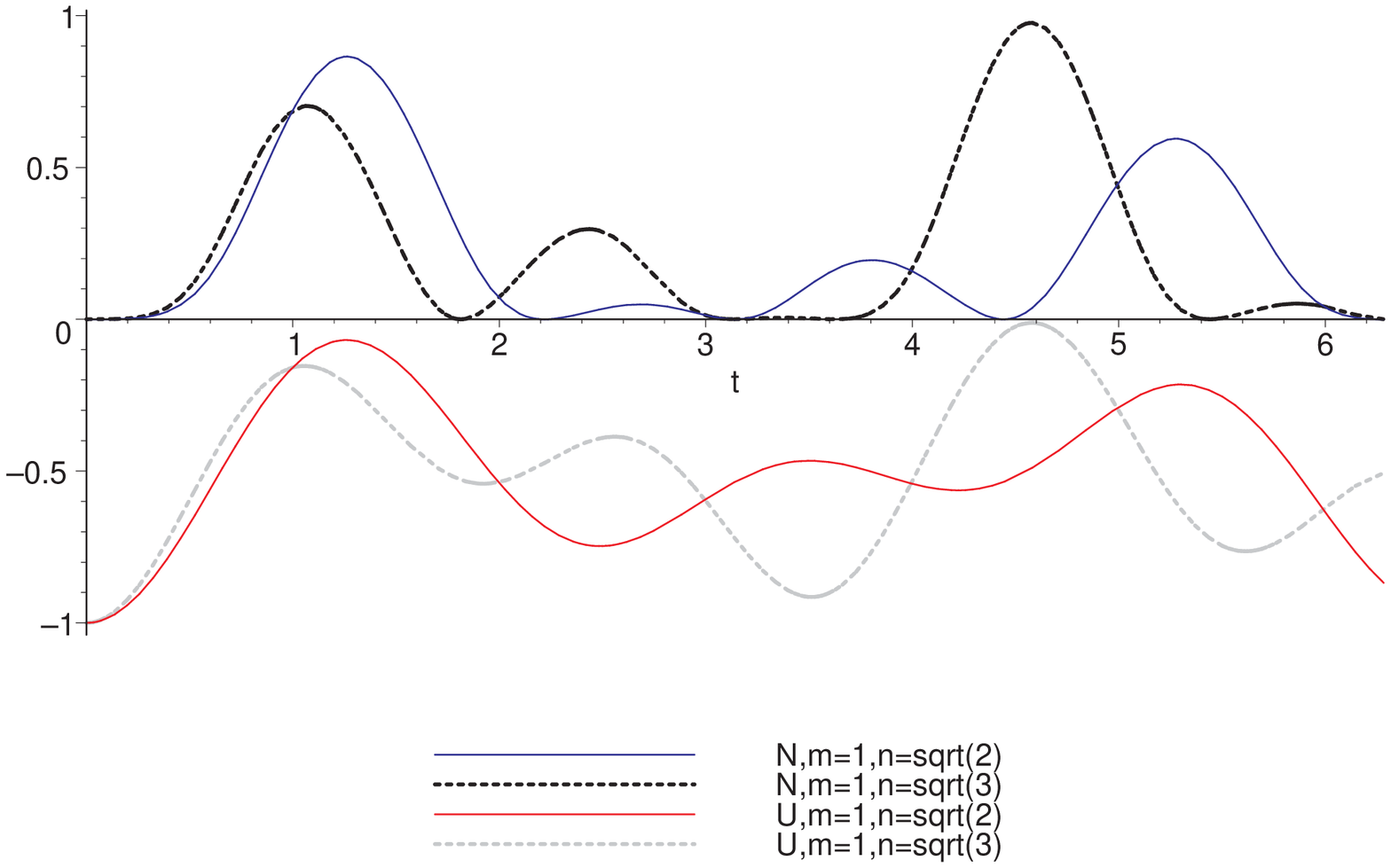}} \\& 
\end{tabular}
\caption{(Color online) Energy $U$ and Negativity $N$ of the state
$\rho_{AB}$ given by Eq.~\eqref{stateAB} \emph{versus} time. The
initial state is given by Eq. \eqref{state0} with $\theta=\pi/4$,
which means a maximally entangled pair $ab$ is initially present,
and its entanglement can be transferred to the pair $AB$. Several
values of $m$ and $n$ were used in the relation $mg_{aA} =
ng_{bB}$.}
 \label{figEnEmT}
\end{figure}
In Fig.~\ref{figEnEmT} we have plotted the temporal behavior of $N$
and $U$ for several values of the coupling constants $g_{aA}$ and
$g_{bB}$. The entanglement transferring process can be followed in
those pictures. The simplest one is for equal coupling constants
$g_{aA} = g_{bB}$, in which the state $\ket{\psi}$ is cyclically
``bouncing'' between the two pairs of qubits. The chosen ratios
between coupling constants indicate a very important behavior of the
system: the complete entanglement transferring process can only
happen if $mg_{aA} = ng_{bB}$ for $m$ and $n$ odd integers.
Eq.~\eqref{aunit} supports this conclusion, since both $g_{aA}t$ and
$g_{bB}t$ must be odd multiples of $\frac{\pi}{2}$ simultaneously
(remember that $\cos ^2\theta$ and $\sin ^2\theta$ are positive
numbers in order to the state $\ket{\psi\de{\theta}}$ to be
entangled). The physical picture is that each pair $aA$ and $bB$
oscillates inside duplets $\ket{10}$ and $\ket{01}$. The situation
is analogous to two classical harmonic oscillators with distinct
frequencies starting from a common extremal point. The implied
relation is necessary for them to meet within an odd number of half
oscillations, which is the condition for a complete transfer of
state. To insist in this point, note that for $g_{aA} = 2g_{bB}$, at
time $t = \frac{\pi}{g_{aA}}$, the pair $aA$ (or, more precisely,
its analogous oscillator) has suffered a full oscillation, but the
pair $bB$ (resp.~ its analogous oscillator) has undergone half an
oscillation, so one could found entanglement between $a$ and $B$,
but no entanglement can be found between any other pair of qubits,
including the studied pair $AB$.

For other values of $\theta$ we obtain similar pictures, with the
only important difference in the entanglement scale, since no
maximally entangled pair will be formed.


\subsection{Entanglement versus energy}

In this subsection we use time as a parameter to draw graphics on an
entanglement versus energy diagram. As we will show, the paths
followed in this phase diagram exhibit interesting patterns.
\begin{figure}[h]\centering
\begin{tabular}{ccc}
    \rotatebox{270}{\includegraphics[scale=0.40]{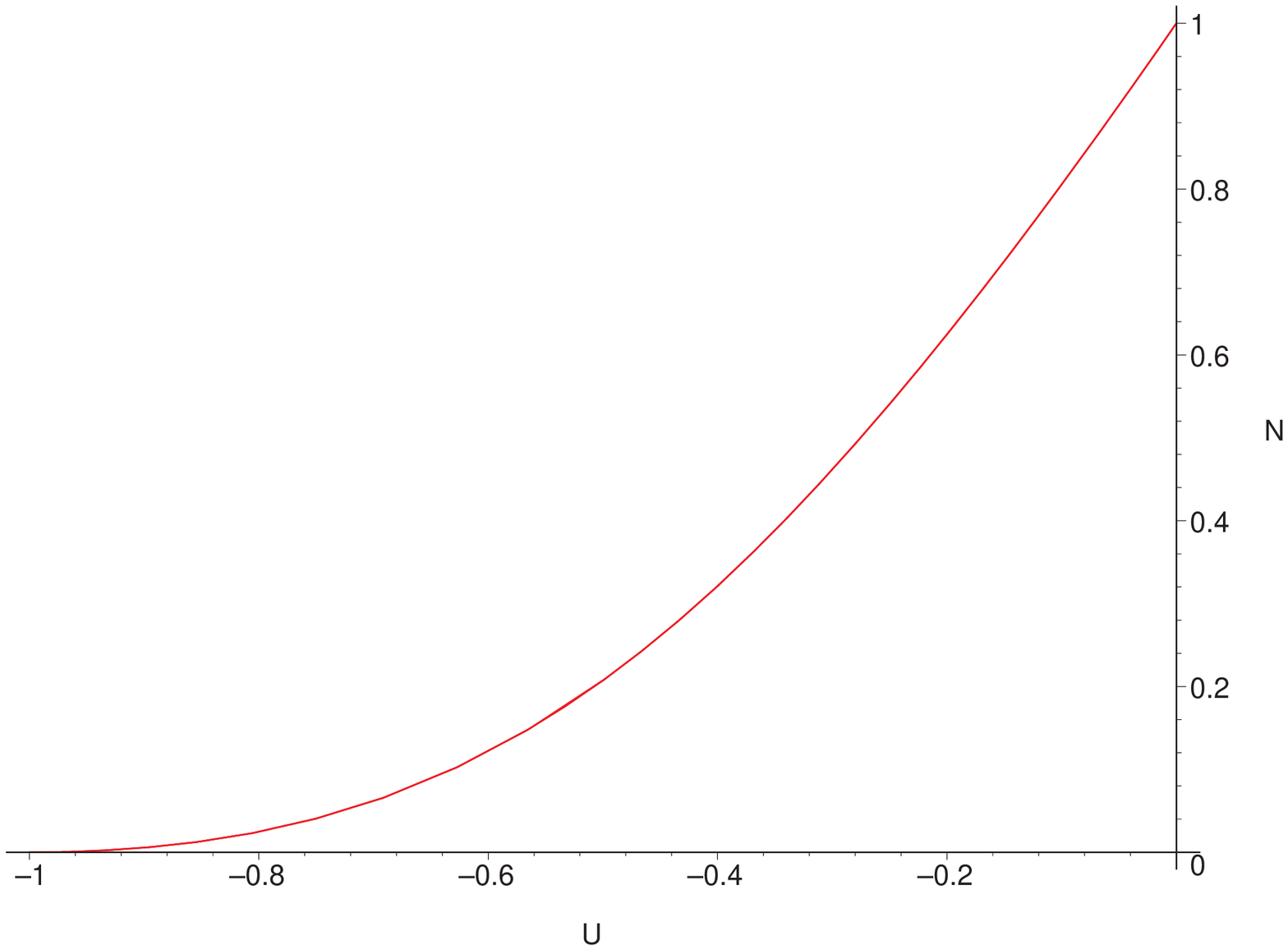}} & \rotatebox{270}{\includegraphics[scale=0.40]{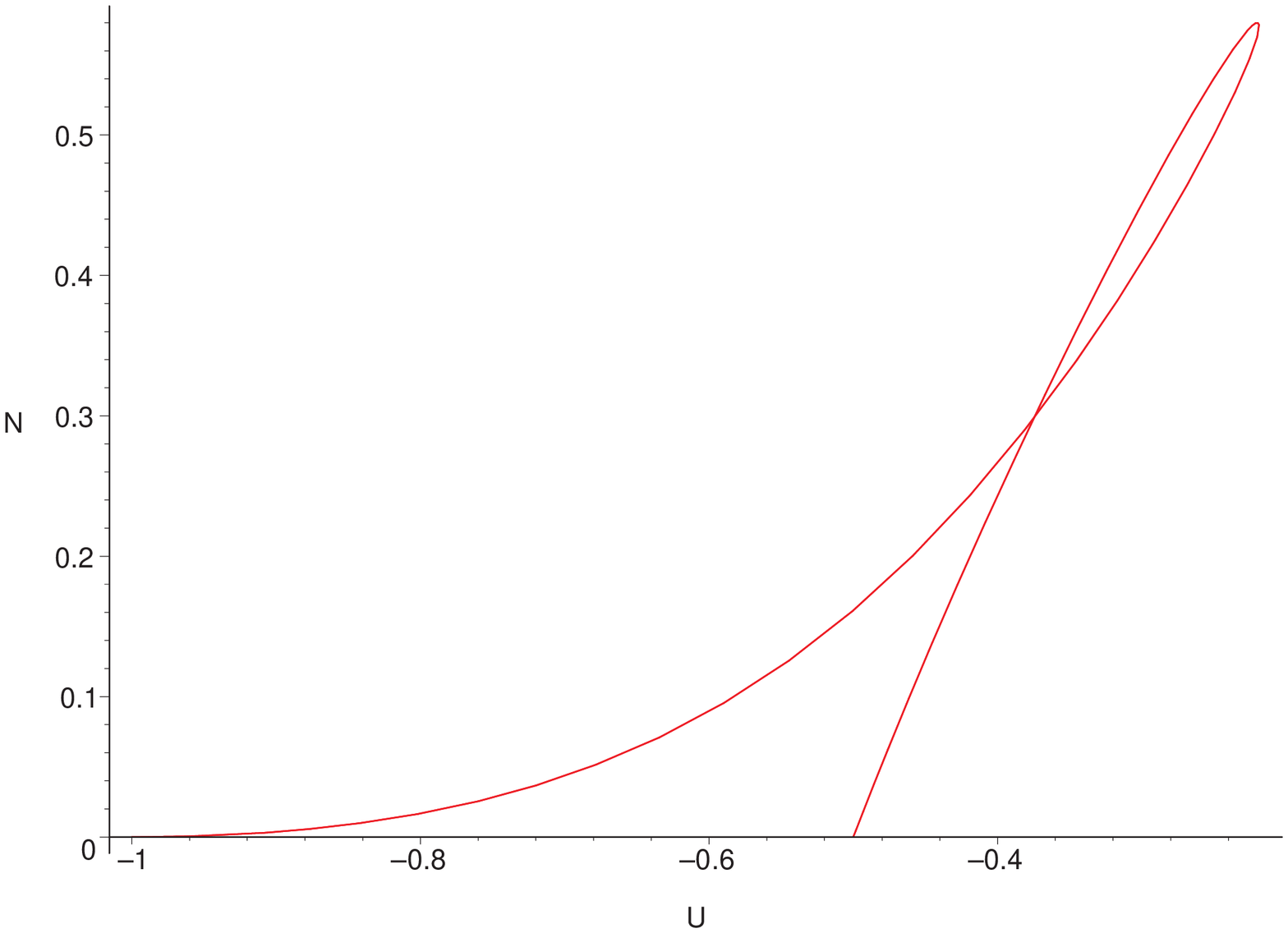}} \\
  \rotatebox{270}{\includegraphics[scale=0.40]{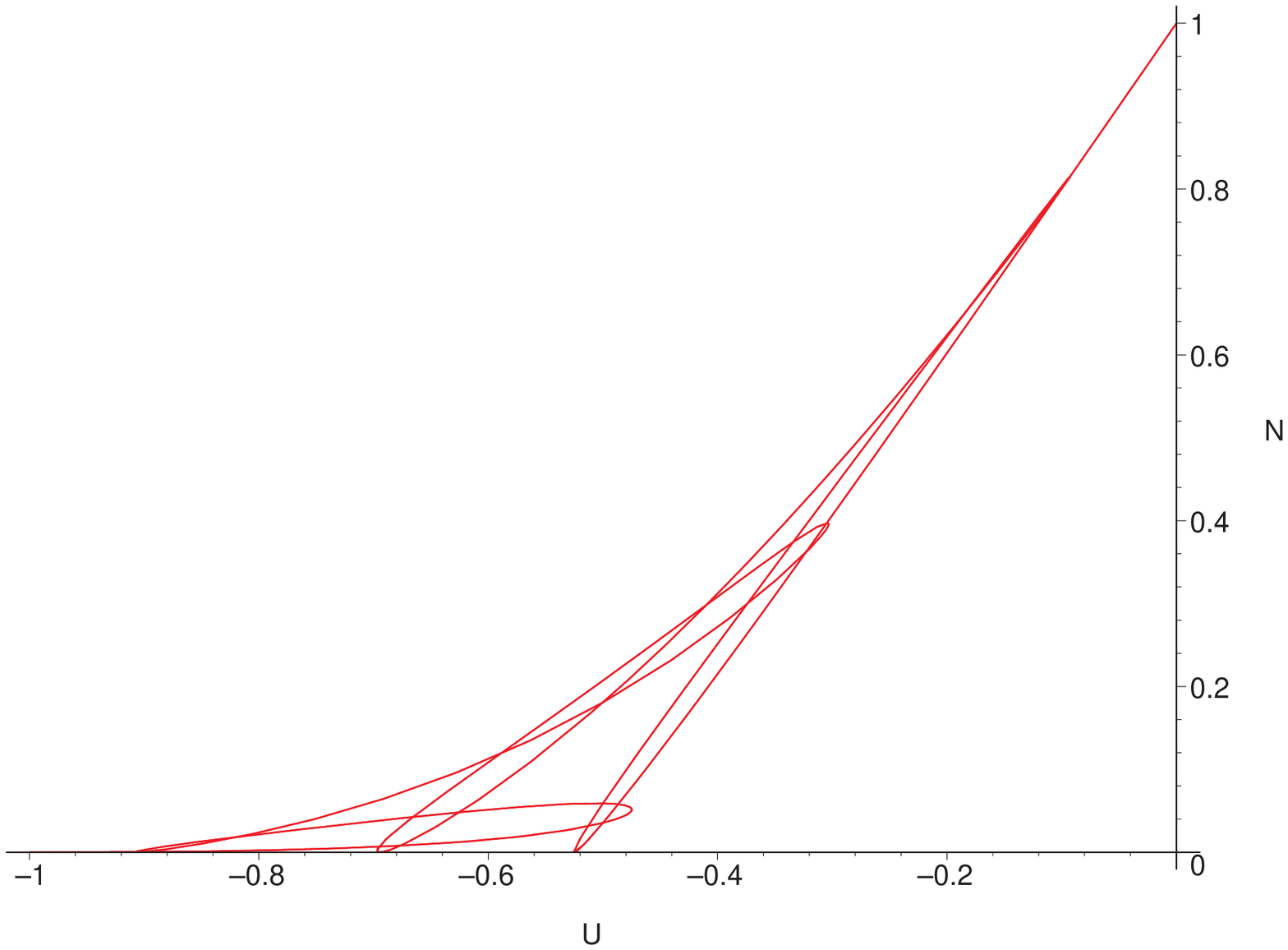}} & \rotatebox{270}{\includegraphics[scale=0.40]{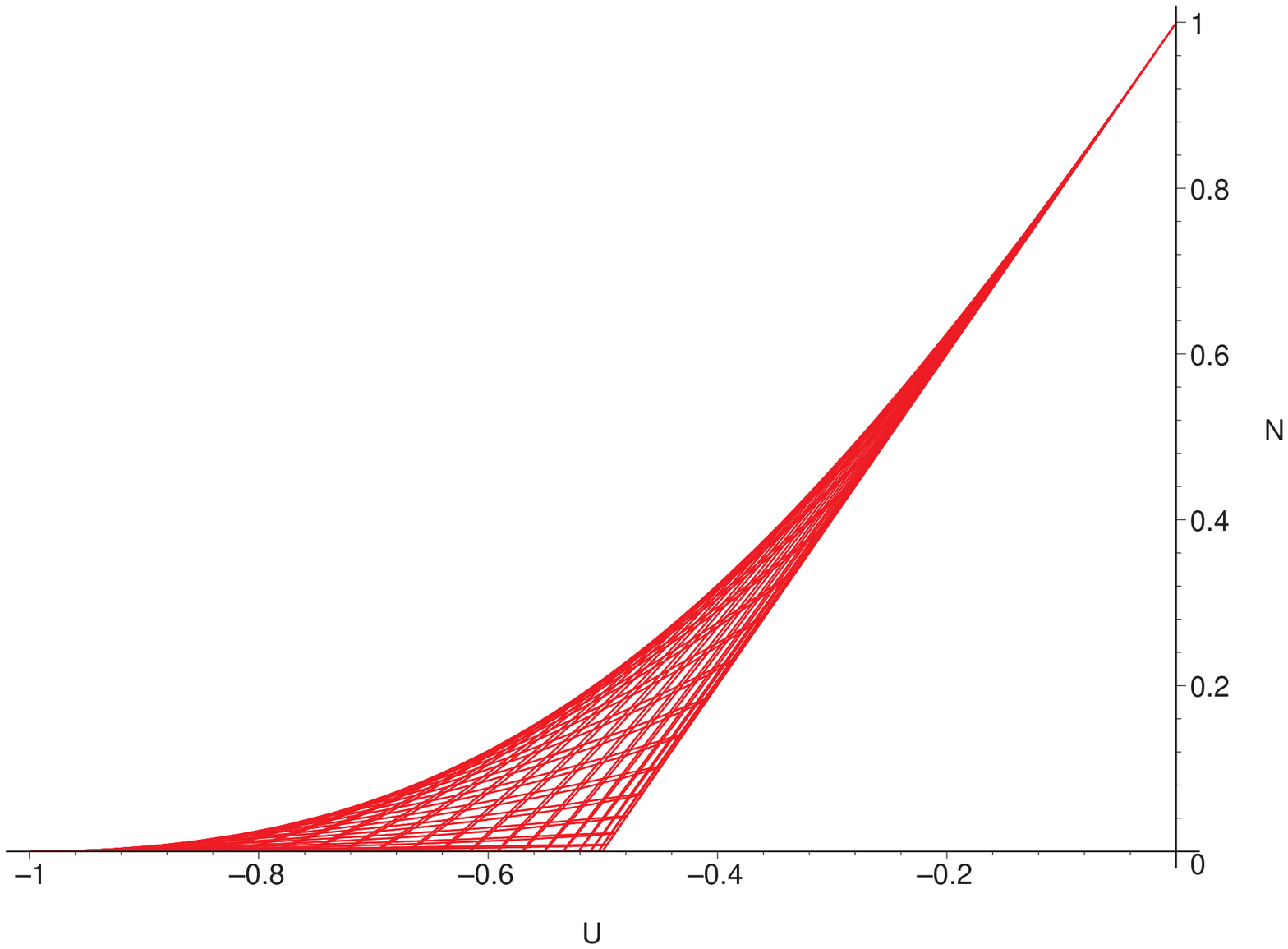}} \\
\end{tabular}
 \caption{(Color online) Negativity ($N$) \emph{versus} Energy ($U$) for $\rho _{AB}$ given by
 Eq.~\eqref{stateAB} with $\theta=\pi/4$, \ie the initial $ab$ state  is maximally entangled.
Parameter relation $g_{aA} =ng_{bB}$ with $n = 1$ (left upper
panel), $n=2$ (right upper panel), $n=7$ (left lower panel), $n=53$
(right lower panel).} \label{EnEmpi4}\end{figure}

\begin{figure}[h]\centering
\begin{tabular}{ccc}
    \rotatebox{270}{\includegraphics[scale=0.40]{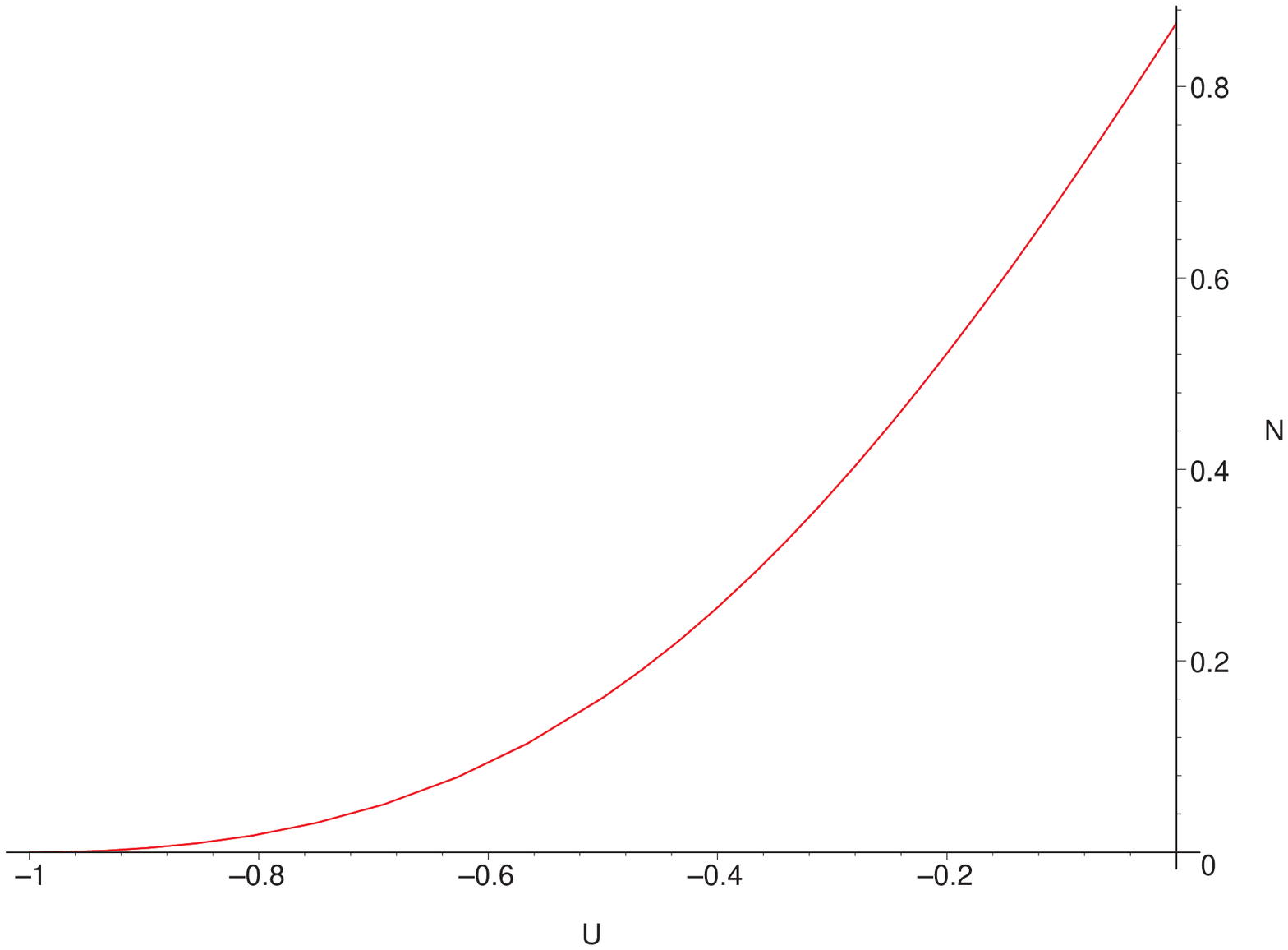}} & \rotatebox{270}{\includegraphics[scale=0.40]{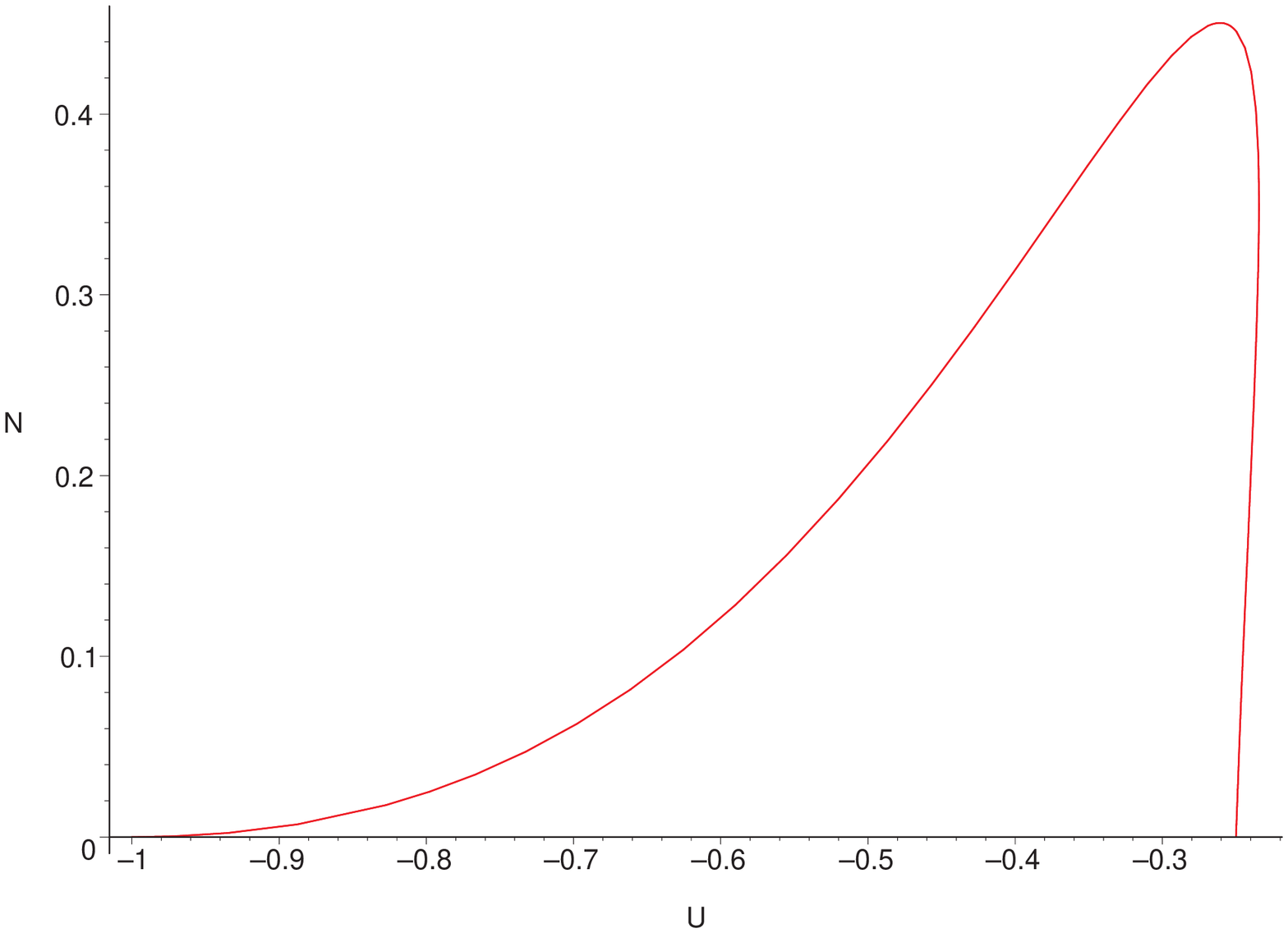}} \\
  \rotatebox{270}{\includegraphics[scale=0.40]{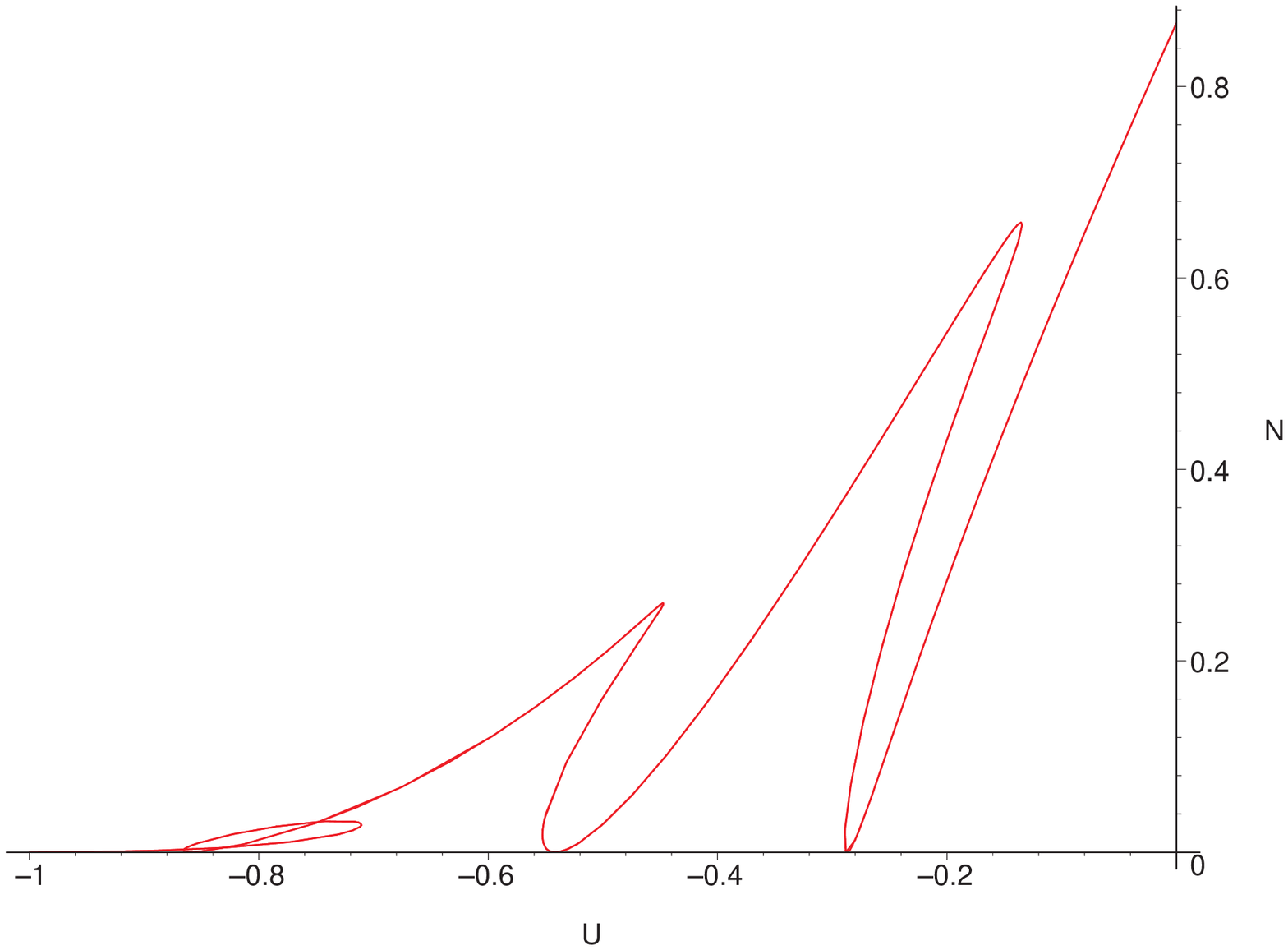}} & \rotatebox{270}{\includegraphics[scale=0.40]{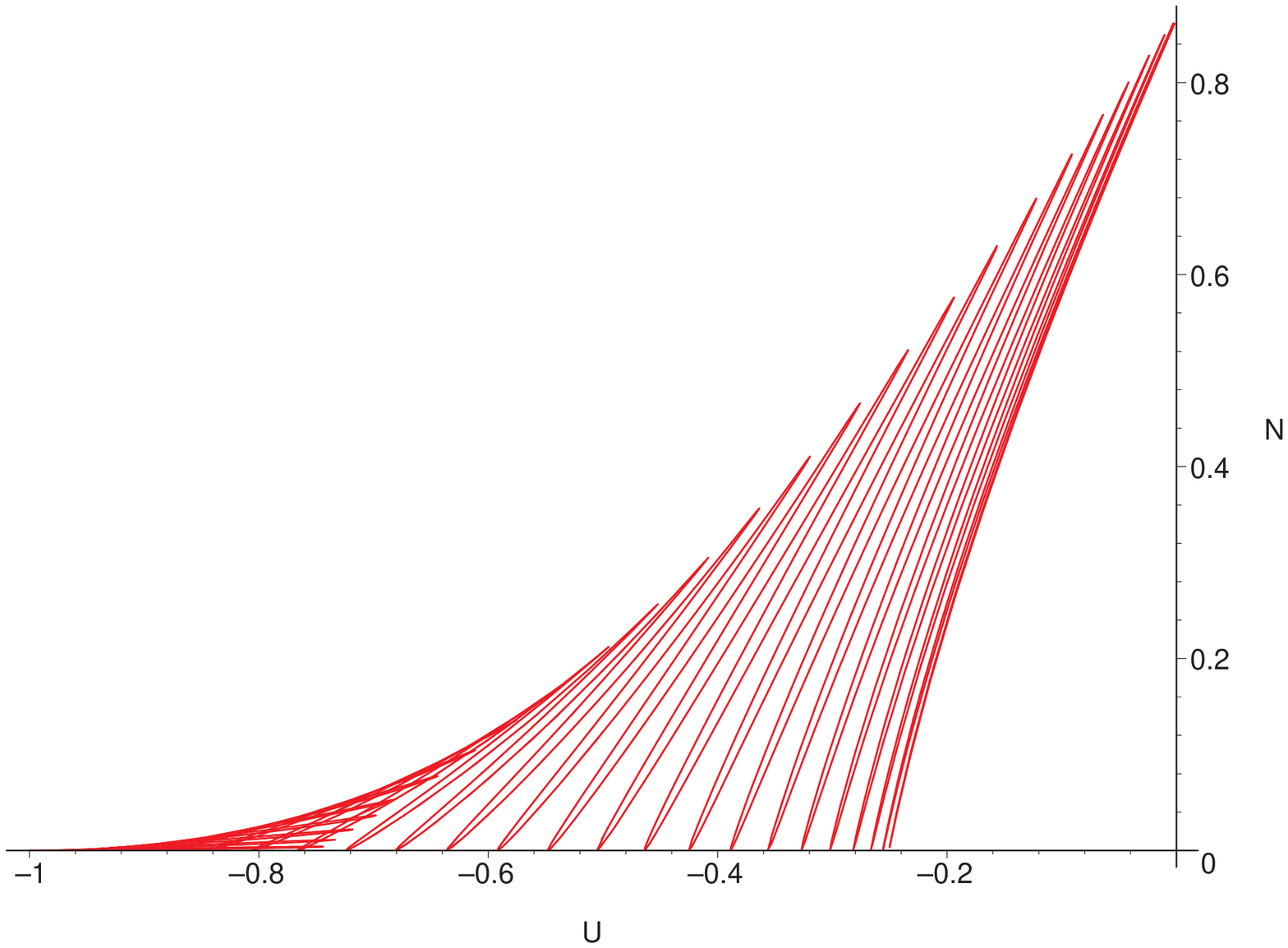}} \\
\end{tabular}
\caption{(Color online) Negativity ($N) $ \emph{versus} Energy ($U$)
for $\rho _{AB}$ given by Eq.~\eqref{stateAB} with $\theta=\pi/3$,
\ie  the initial $ab$ is \emph{partially} entangled. Parameter
relation $g_{aA} =ng_{bB}$ with $n = 1$ (left upper panel), $n=2$
(right upper panel), $n=7$ (Left lower panel), $n=53$ (right lower
panel).} \label{EnEmpi3}\end{figure}

\begin{figure}[h]\centering
\rotatebox{270}{\includegraphics[scale=0.40]{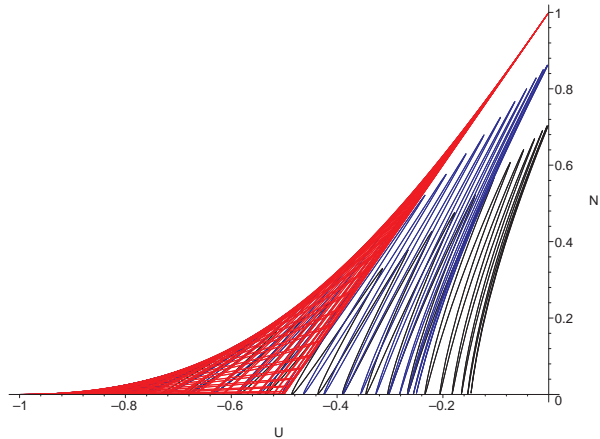}} \\
\caption{(Color online) Negativity ($N$) \emph{vs.} Energy ($U$) for
$\rho_{AB}$ given by Eq.~\eqref{stateAB} with $\theta=\pi/4$ (red),
$\theta=\pi/6$ (blue), and $\theta=\pi/8$ (black). Parameter
relation $g_{aA}=53g_{bB}$.} \label{EnEmtheta}
\end{figure}

The graphics for $\theta=\pi/4$ (Fig.~\ref{EnEmpi4}) and
$\theta=\pi/3$ (Fig.~\ref{EnEmpi3}) are qualitatively different.
However, it can be seen through Fig.~\ref{EnEmtheta} that,
independent of the available initial entanglement (given by the
value of $\theta$), the accessible region in the parameter space $N
\times U$ is bounded by an upper curve. This bound can be explained
in the following way: using the fact that $U=-a$ in
Eq.~\eqref{negat} we have
\begin{equation}\label{genRel}
N^2-2NU=4|d|^2.
\end{equation}
However, when $g_{aA}=g_{bB}$ and $\theta=\pi/4$ we have $b=c=d$, in
which case $N^2-2NU=4b^2$. Using the normalization condition
$a+b+c=1$, we get
\begin{equation}
(1+U)^2=4d^2.
\end{equation}
Therefore, in the ideal situation of equal coupling
$g_{aA}=g_{bB}$ and maximally entangled initial $ab$ state
($\theta=\pi/4$), we can write
\begin{equation}\label{curvaotima}
N^2-2NU=(1+U)^2.
\end{equation}

As we will see now this equation is exactly the one that limits the
phase-space for quantum states in this problem. The normalization
condition yields $b+c=1+U$ that allows us to obtain
\[
4bc=(U+1)^2-(b-c)^2,
\]
which implies
\begin{equation}
4bc\leq(U+1)^2.
\end{equation}
At the same time, the condition for matrix \eqref{stateAB} to be
considered a true density matrix is that its eigenvalues are all
positive, which is reached if and only if $|d|^2\leq bc$. Therefore,
we can conclude that
\begin{equation}
4|d|^2\leq4bc\leq(U+1)^2.
\end{equation}
and, from Eq. \eqref{genRel}, we find
\begin{equation}
N^2-2NU\leq(U+1)^2.
\end{equation}
As we saw in Eq.~\eqref{curvaotima} the equality is reached for
$g_{aA}=g_{bB}$ and $\theta=\pi/4$. So Eq.~\eqref{curvaotima} bounds
the region that density matrices of the form \eqref{stateAB} can
occupy in the diagram $N\times U$ .

\section{Open system}\label{OpSys}
Up to this moment we have considered any two pairs of qubits. Now we
will adhere to one specific physical realization, namely: two atoms
resonantly coupled to two independent cavity modes. If the cavities
are initially in the ground state (vacuum, no photon) and the usual
approximations are valid~\cite{MandelWolf}, the Jaynes-Cummings
Hamiltonian
\begin{equation}
H_{JC} = \frac{\hbar \nu}{2} \sigma _z + \hbar \omega
\de{a^{\dagger}a + \frac12} + \hbar \Omega \de{a^{\dagger}\sigma _-
+ a\sigma _+}
\end{equation}
essentially reduces to the form \eqref{HxX}, with the lower case
qubit representing the two-level atom and the capital one, the first
two energy levels of the field mode. Hence, the situation here
studied models an experiment where previously entangled atoms
transfer such entanglement to independent cavity modes.
%
%
One nice point when considering this particular situation of atoms
transferring entanglement and energy to resonant cavities is that
those systems can have very different dissipation times. In fact,
atomic levels can be selected so that their dissipation time scale
is much larger than those of typical resonant cavities. In this
case, one can ask what happens if the system that receives the
energy and the entanglement dissipates it to an external reservoir.
In order to answer this question, the unitary analysis considered up
to now has to be abandoned and we must change from a Hamiltonian
approach to a master equation one.

We will consider that the qubits $AB$, now represented by the cavity
field modes, are in contact with independent reservoirs and interact
with them. Since atomic lifetimes (for the atomic transitions used
in cavity QED experiments) are usually much greater than cavity
decay times, we will not couple the lower case qubits to any
external device.
%

To address this problem we consider the time evolution of the global
system described by the master equation in the Lindblad form
\cite{Lindblad}:
\begin{equation}\label{eqmestra}
    \frac{d}{dt}{\rho}_{aAbB}(t)=\frac{1}{i\hbar}\bigl[H,\rho_{aAbB}
    (t)\bigr]+\frac{1}{2\hbar}\sum_{i}\Bigl(\bigl[\hat{V}_{i}\rho_{aAbB}(t),\hat{V}
    _{i}^{\dag}\bigr]+\bigl[\hat{V}_{i},\rho_{aAbB}(t)\hat{V}_{i}^{\dag}\bigr]\Bigr)
    \ ,
\end{equation}
where $H$ is given by \eqref{hamilt1} and the operators
$\hat{V}_{i}$ and $\hat{V}_{i}^{\dag}$ describe the effects of the
coupling to the reservoirs. For simplicity, we will model only the
dissipation of energy in the cavities coupled to null temperature
reservoirs, which can be done using
\begin{subequations}
\begin{eqnarray}
\hat{V}_{1} &=& \sqrt{2\hbar\kappa_{A}}
{\sigma}_{-}^{A}\otimes I_{abB} \label{V1}\\
  \hat{V}_{2} &=& \sqrt{2\hbar\kappa_{B}} \ I_{aAb}\otimes
  {\sigma}_{-}^{B} \label{V2}\\
  \hat{V}_{i} &=& 0 \ , \ \forall \ i > 2 \label{Vi}
\end{eqnarray}
\end{subequations}
The constants $\kappa_{A}$ and $\kappa_{B}$  are directly given by
the decay rates of each cavity mode.

\begin{figure}[h]\centering
\begin{tabular}{cc}
\rotatebox{270}{\includegraphics[scale=0.45]{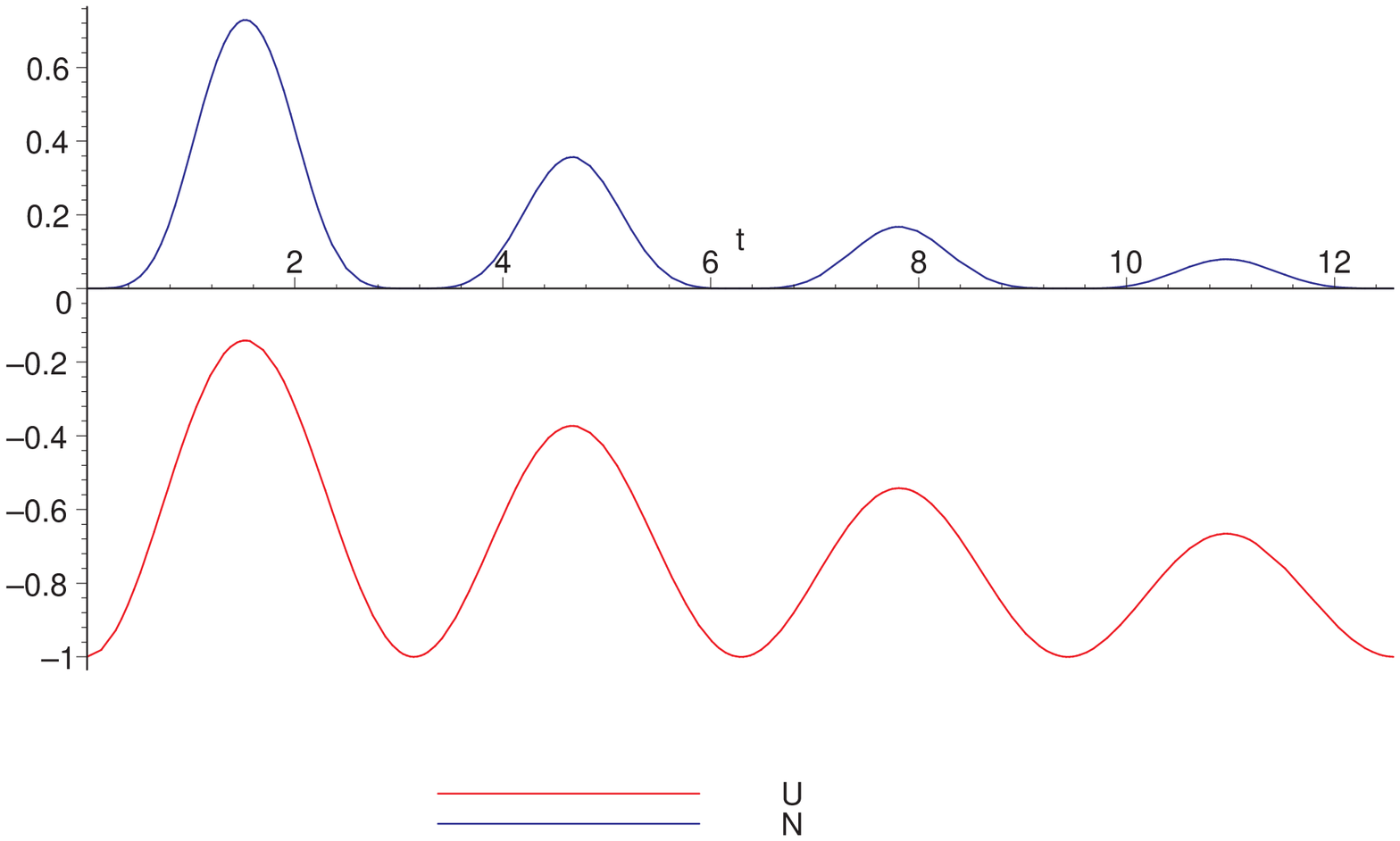}} & \rotatebox{270}{\includegraphics[scale=0.45]{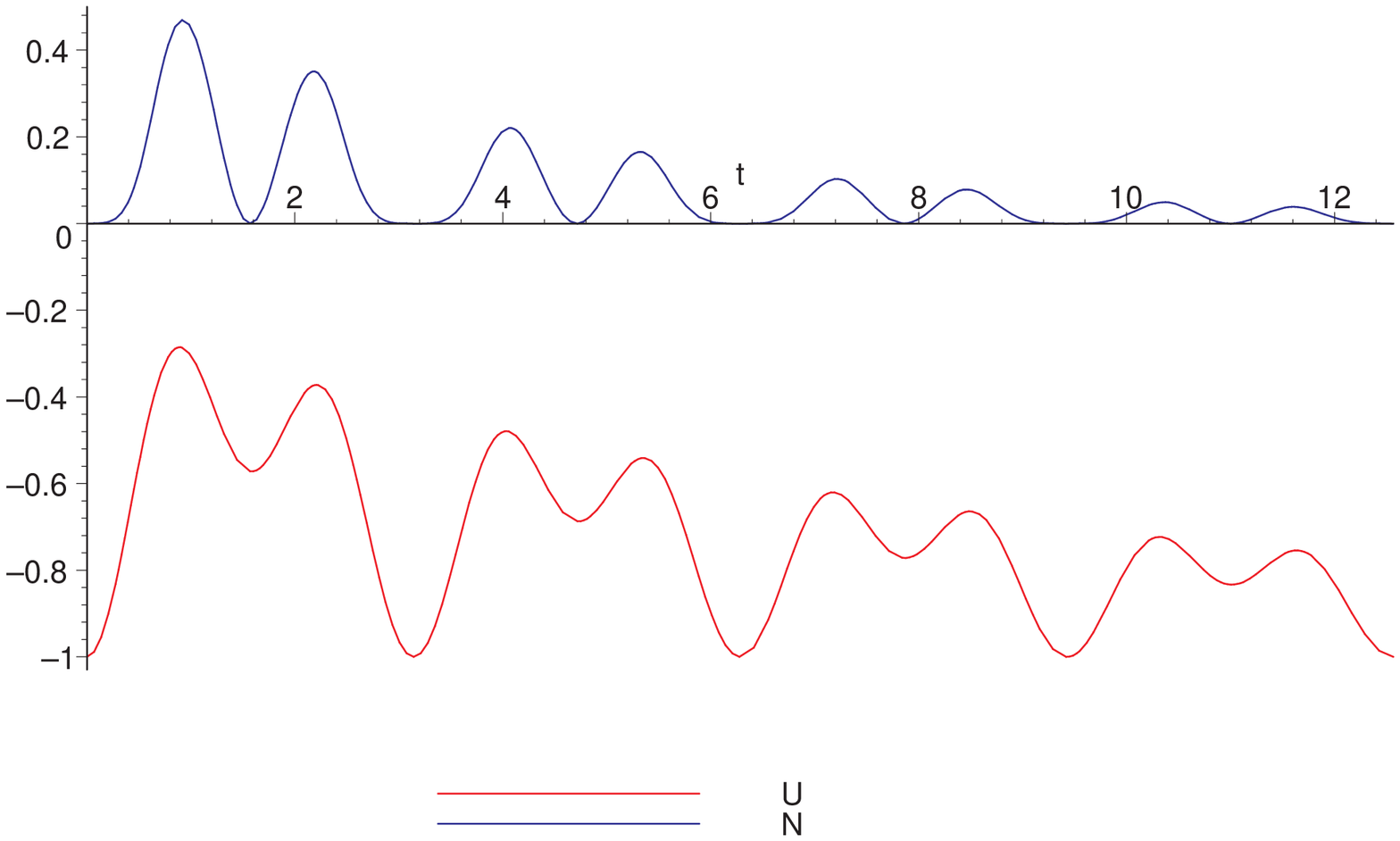}} \\
\rotatebox{270}{\includegraphics[scale=0.45]{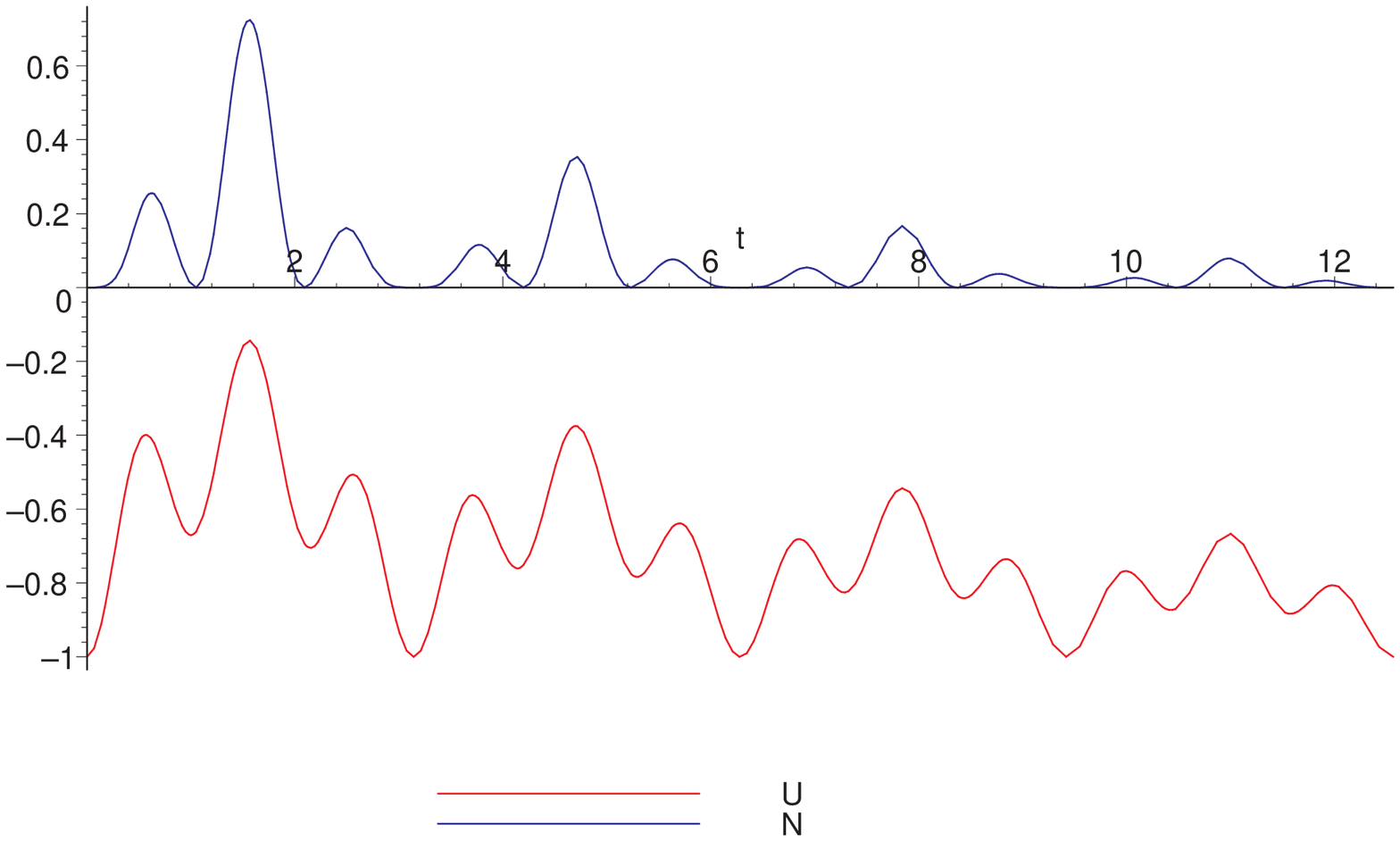}} & \rotatebox{270}{\includegraphics[scale=0.45]{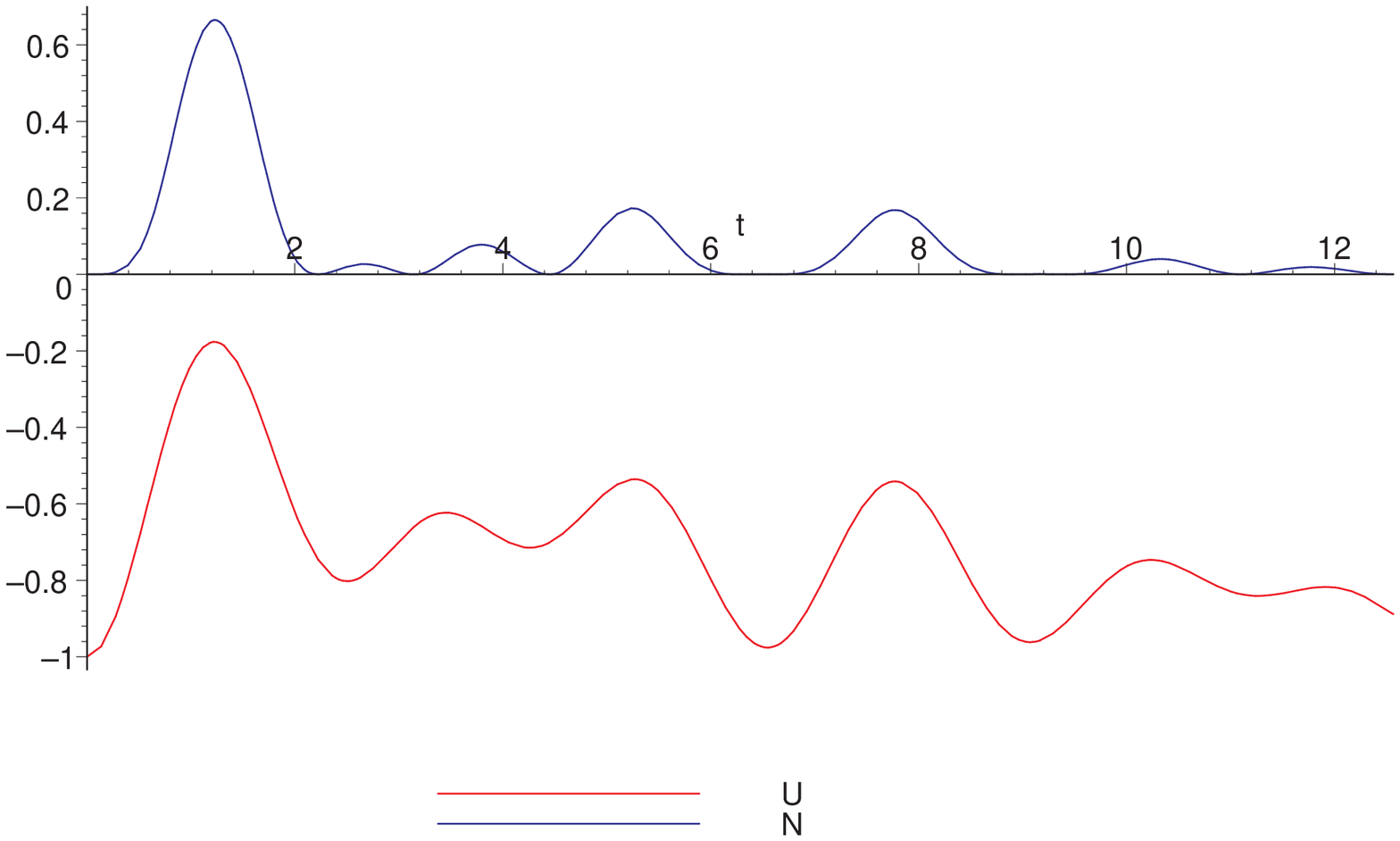}} \\
\end{tabular}
\caption{(Color online) Energy $U$ (red) and Negativity $N$ (blue)
of the cavity modes state \eqref{stateAB} with matrix elements given
by Eq.~\eqref{openelements} \emph{versus} time. The initial state of
the system is given by \eqref{state0} with $\theta=\pi/4$, \ie, a
maximally entangled $ab$ state. Parameter relations:
$\kappa_{aA}=\kappa_{bB}=0.1 g_{aA}$ and $g_{bB}/g_{aA}=n$.
\emph{Left-above:} $n=1$. \emph{Right-above:} $n=2$.
\emph{Left-below:} $n=3$. \emph{Right-below:} $n=\sqrt{2}$.}
  \label{UNvsTopen}
\end{figure}

Starting from the special case of the initial state \eqref{state0},
given by $\theta = \frac{\pi}{4}$ (a Bell state for the donor pair
of qubits), after a somewhat lengthy calculation we find that the
state of both cavities can still be written in the form of
Eq.~\eqref{stateAB}, but now with the matrix elements given by
\begin{subequations}
\begin{eqnarray}
a&=&1-\Biggl\{\biggl[\frac{\sqrt{2}\,\,g_{aA}}{\Omega_{aA}}
\sin\Bigl(\frac{\Omega_{aA}}{2}t\Bigr)e^{-\kappa_{A}t/2}\biggr]^{2}+
\biggl[\frac{\sqrt{2}\,\,g_{bB}}{\Omega_{bB}}\label{eq50}
\sin\Bigl(\frac{\Omega_{bB}}{2}t\Bigr)e^{-\kappa_{B}t/2}\biggr]^{2}\Biggr\}\\
b&=&\biggl[\frac{\sqrt{2}\,\,g_{bB}}{\Omega_{bB}}\label{eq49}
\sin\Bigl(\frac{\Omega_{bB}}{2}t\Bigr)e^{-\kappa_{B}t/2}\biggr]^{2}\\
c&=&\biggl[\dfrac{\sqrt{2}\,\,g_{aA}}{\Omega_{aA}}\label{eq47}
\sin\Bigl(\dfrac{\Omega_{aA}}{2}t\Bigr)e^{-\kappa_{A}t/2}\biggr]^{2}\\
d&=&\biggl[\dfrac{\sqrt{2}\,\,g_{aA}}{\Omega_{aA}}\label{eq48}
\sin\Bigl(\dfrac{\Omega_{aA}}{2}t\Bigr)e^{-\kappa_{A}t/2}\biggr]\cdot
\biggl[\dfrac{\sqrt{2}\,\,g_{bB}}{\Omega_{bB}}\sin\Bigl(\dfrac{\Omega_{bB}}{2}t\Bigr)
e^{-\kappa_{B}t/2}\biggr],
\end{eqnarray}
\label{openelements}
\end{subequations}
with the definitions
\begin{subequations}
\begin{eqnarray}
\Omega_{aA}&=& \sqrt{4g_{aA}^{2}-\kappa^{2}_{A}},\label{eq51} \\
\Omega_{bB}&=& \sqrt{4g_{bB}^{2}-\kappa^{2}_{B}}.\label{eq52}
\end{eqnarray}
\end{subequations}
Since the state of the system $AB$ is still described by density
matrices of the form~\eqref{stateAB} we expect the existence of the
same bounds in the energy-time diagram~\eqref{curvaotima}. Also note
that energy and negativity are still respectively described by
equations of the forms~\eqref{GenU} and \eqref{negat}.
In Fig. \ref{UNvsTopen} we have plotted some curves of negativity
and energy vs. time, for the system $AB$ considering the temporal
evolution of the matrix elements given by Eq. \eqref{openelements}.
Note that the graphics are qualitatively similar to the ones
displayed in Fig.~\ref{figEnEmT} for unitary evolutions. However,
as expected, both entanglement and energy decay exponentially to
their lowest values as a function of time. This can be understood
from the fact that the environment drives the system exponentially to
the state $\ket{00}$ (see the matrix elements in Eq.
\eqref{openelements} - the element \eqref{eq50} goes to unity while
all the others go to zero), which has no entanglement and has the
minimum energy value $U=-1$.

\begin{figure}[h]\centering
\begin{tabular}{cc}
\rotatebox{270}{\includegraphics[scale=0.45]{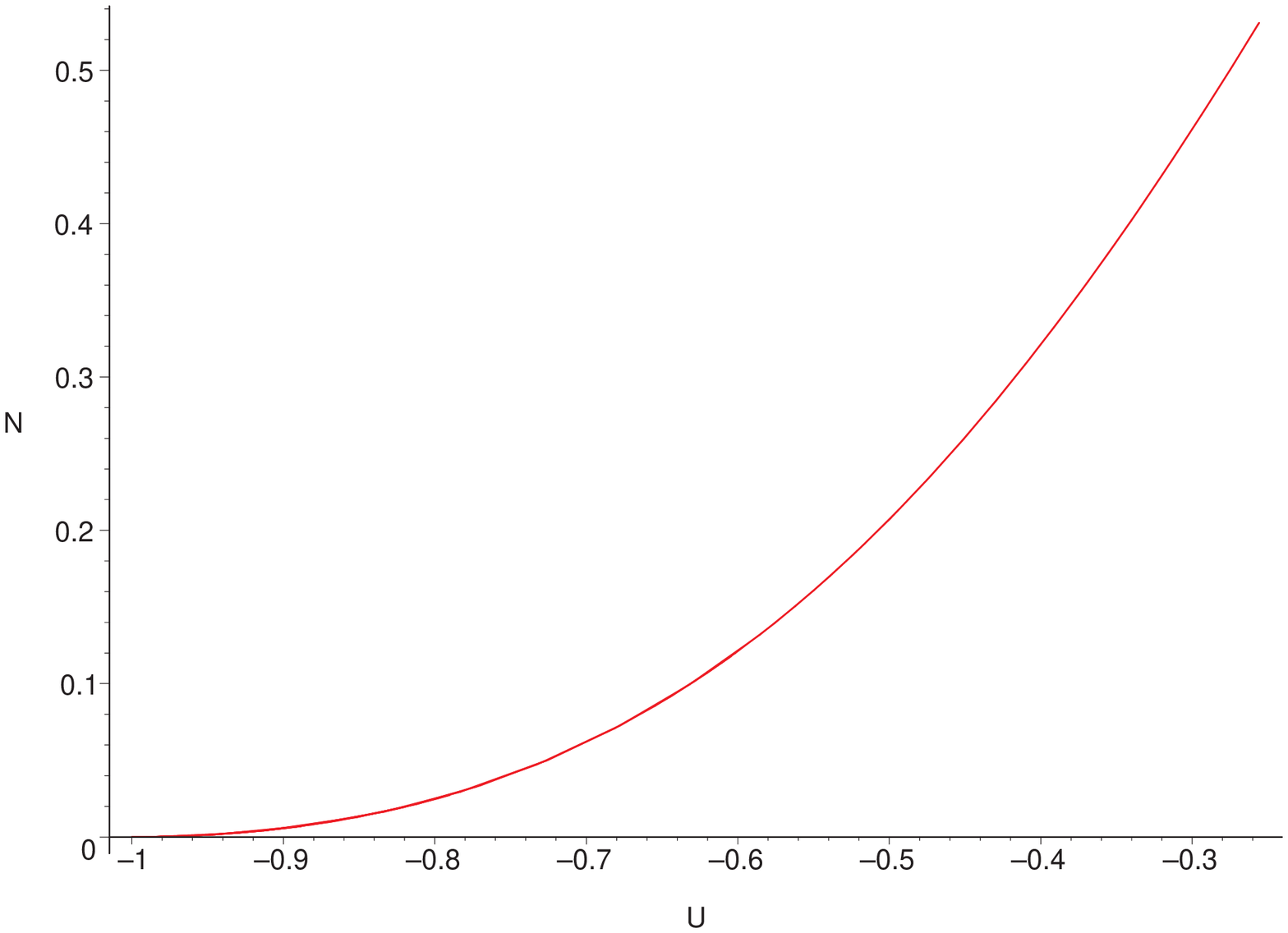}} & \rotatebox{270}{\includegraphics[scale=0.45]{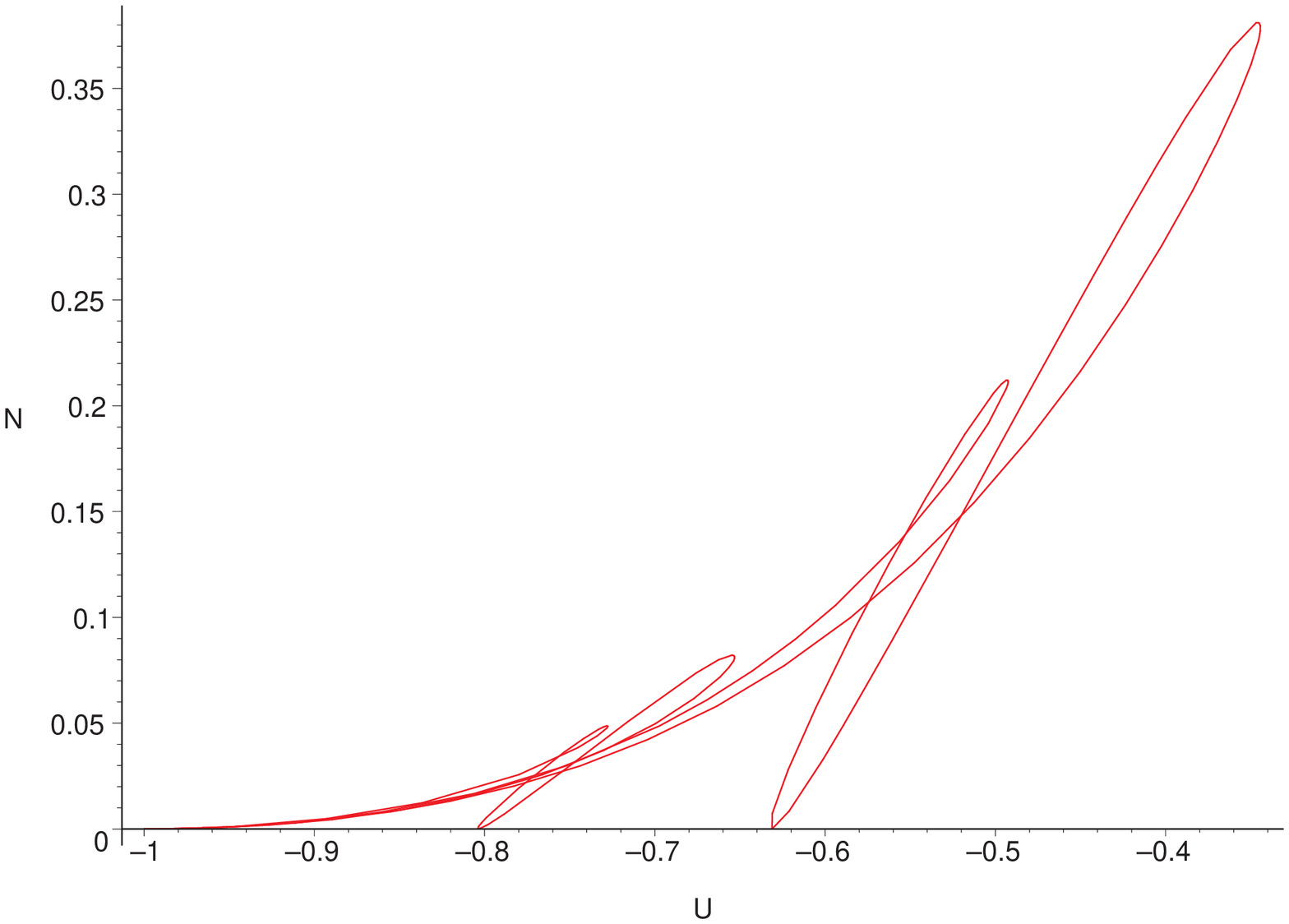}} \\
\rotatebox{270}{\includegraphics[scale=0.45]{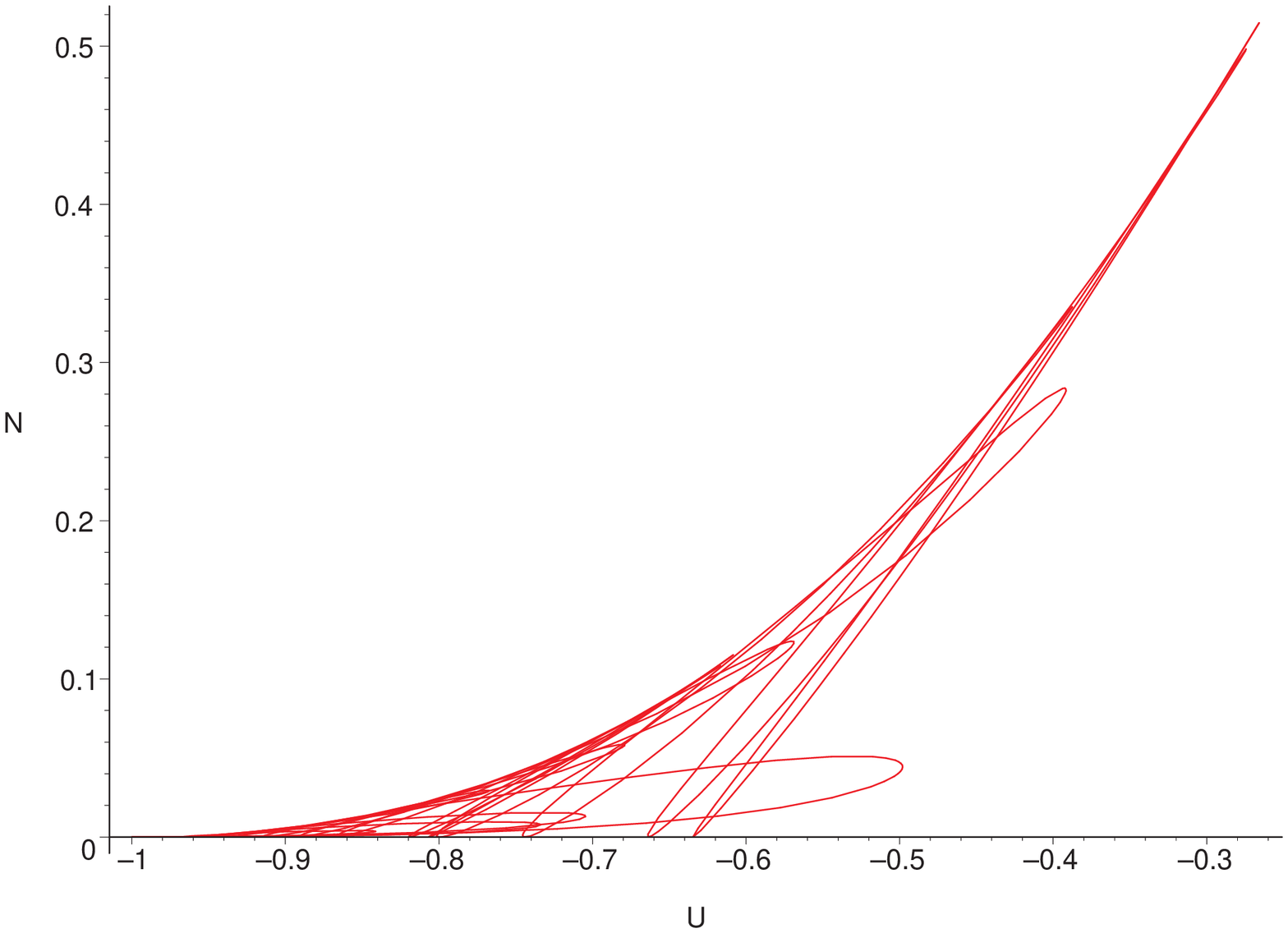}} & \rotatebox{270}{\includegraphics[scale=0.45]{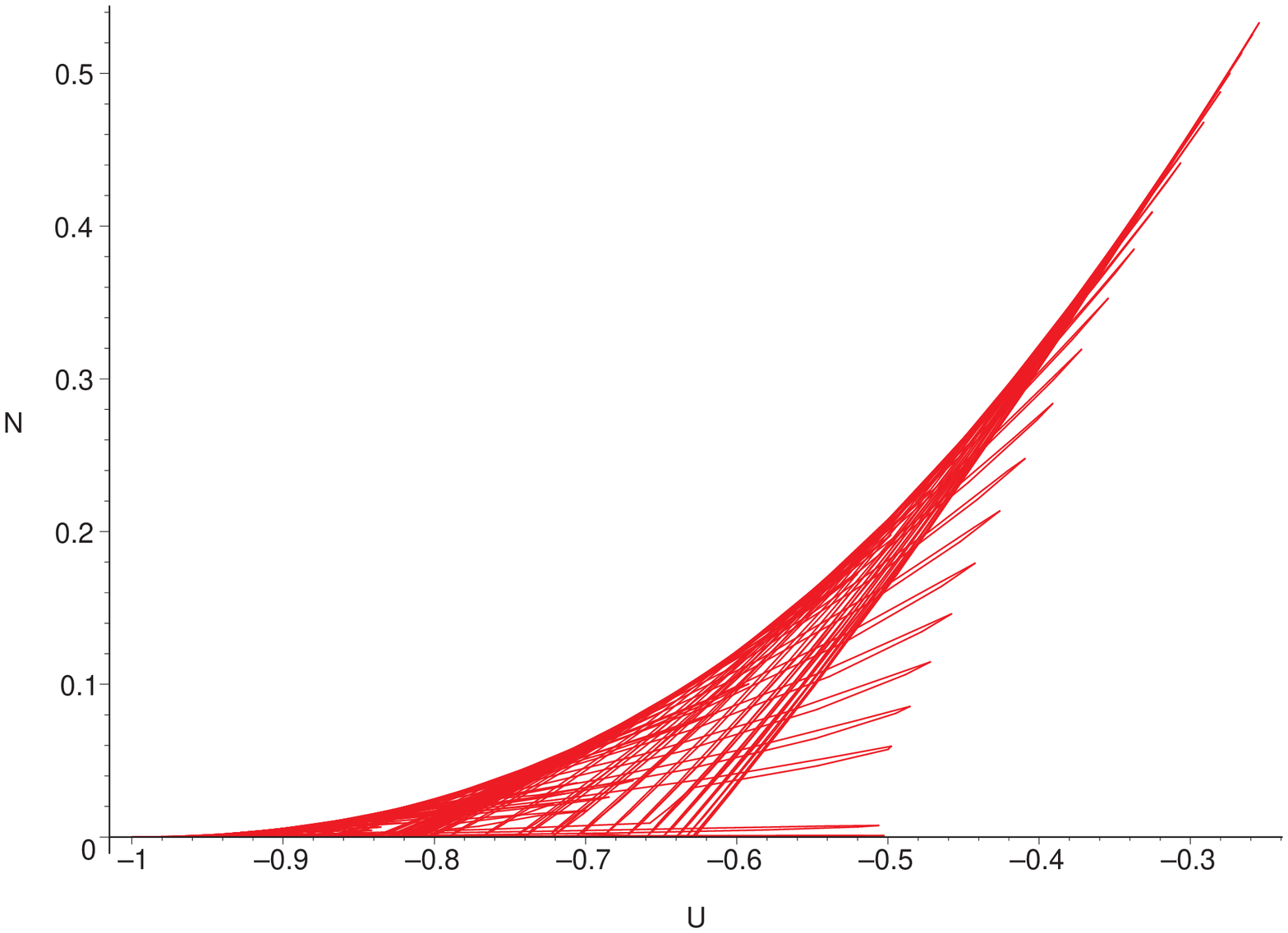}} \\
\end{tabular}
\caption{(Color online) Negativity ($N$) $vs.$ Energy ($U$) of the
cavity modes state \eqref{stateAB} with matrix elements given by
Eq.~\eqref{openelements} for fixed decay rates and different
couplings. The initial state of the system $ab$ is given by
\eqref{state0} with $\theta=\pi/4$. Parameter relations:
$\kappa_{aA}=\kappa_{bB}=0.1 g_{aA}$ and $g_{aA} =n g_{bB}$ with $n
= 1$ (left upper panel), $n=2$ (right upper panel), $n=7$ (Left
lower panel), $n=53$ (right lower panel).}
 \label{UNopen2}
\end{figure}

We have also plotted the negativity vs. energy for the non-unitary
case for different values of $g_{aA}$, $g_{bB}$, $\kappa_{aA}$ and
$\kappa_{bB}$. This is displayed in Fig. \ref{UNopen2}. As commented
before the followed paths are still bounded by the same limits of
the unitary case. However, as the dissipative mechanisms get
stronger (i.e. coefficients $\kappa_{A}$ and $\kappa_B$ get closer
to the coupling constants $g_{aA}$ and $g_{bB}$), less and less
entanglement is transferred to subsystem $\rho_{AB}$.

A different phenomenology can be anticipated for the case of
non-null temperature. Since thermal photons can now be captured by
both cavities, the state $\ket{11}$ will be populated and the form
\eqref{stateAB} will not be valid anymore. One consequence is that
one can expect the phenomenon known as entanglement sudden
death~\cite{SD}, since it will not be necessary to nullify the
elements $d$ in order to have a positive partial transpose, and
hence no entanglement.

\section{Discussions}\label{Conc}

In this paper we have addressed the problem of entanglement and
energy transfer between pairs of qubits. We considered the
particular example of two atoms interacting with two cavities in the
Jaynes-Cummings model. This evolution can be seen as a state
transferring process and, for specific coupling constants, a
dynamical entanglement swapping. If the atoms are initially in an
entangled state, this entanglement is fully or partially transferred
to the cavities depending on coupling constants and time. This
entanglement swapping process is accompanied by an energy transfer
as well, and we have shown that entanglement and energy in the
cavities system are strictly related.

To clarify this relation we studied these quantities in various
scenarios. First we considered the whole system as isolated, and
investigated its time evolution for several coupling constants and
different initial atomic entanglement. In each case, we traced out
the atoms (as we now refer to the lower case qubits), we drew an
{entanglement vs. energy} phase-diagram for the cavity modes and we
found an upper-limit for all the possible paths in these diagrams.
This bound corresponds to maximally entangled atoms transferring its
entanglement and energy to independent cavities at exactly the same
rates.

We also considered the possibility of dissipation in the cavities.
In this case, while the atoms are transferring excitations and
entanglement to the cavities, some energy is lost to the
environment. The cavities state goes asymptotically to state
$\ket{00}$. In the dissipative regime, the evolution of entanglement
and energy of the cavities state exhibits the same characteristics
pointed out for the unitary case, \ie, the paths followed in the
entanglement-energy diagram are limited to a restricted region whose
frontier is identified by the trajectory described when the
couplings are identical and the initial entanglement of the atomic
state is maximum. However, as expected, neither entanglement nor
energy can be fully transferred unless the dissipative times are
much larger than the inverse Rabi frequencies involved. Fortunately,
this regime is usually achieved in cavity QED experiments.

We only analyzed the entanglement between qubits $AB$ (the modes, in
the physical realization proposed). However, most of the time the
whole system presents multipartite entanglement which may provide
interesting new results if further studied. Other important
continuations of this work include the treatment of non-completely
resonant systems (\eg  $\omega _a = \omega _A \neq \omega _b =
\omega _B$, which corresponds to two distinct atoms resonantly
coupled to cavity modes, as well as the case of dispersive coupling)
and also other couplings to reservoirs, like including temperature
in the scenario here presented and also considering spontaneous
decay for the atoms.

\begin{acknowledgments}
The authors recognize fruitful discussions with M.C. Nemes.
Financial support from CNPq and PRPq-UFMG is acknowledged. This work
is part of the Milleniun Institute for Quantum Information project
(CNPq).
\end{acknowledgments}

\end{document}